\documentclass[journal,onecolumn]{IEEEtran}
\normalsize
 \usepackage[T1]{fontenc}% optional T1 font encoding
\usepackage[linesnumbered,ruled]{algorithm2e}
\usepackage{stfloats,color}
\usepackage{amsfonts}
\usepackage{amssymb}
\usepackage[cmex10]{amsmath}
\usepackage{graphicx}
\usepackage{setspace}
\usepackage{cite}
\usepackage{array}
\usepackage{subcaption}
\usepackage{times}
\usepackage{epsfig}
\usepackage{latexsym}
\usepackage{epstopdf}
\usepackage{verbatim}
\usepackage{units}
\usepackage{amsthm}
\usepackage{placeins}
\usepackage{afterpage}
\usepackage{dsfont}
\usepackage{soul}
\usepackage{multicol}
\usepackage{multirow}
\usepackage{mathtools}
\usepackage[cmintegrals]{newtxmath}
\newcolumntype{P}[1]{>{\centering\arraybackslash}p{#1}}
\newcolumntype{M}[1]{>{\centering\arraybackslash}m{#1}}

\newcommand{\defeq}{\ensuremath{\triangleq}}

\usepackage[colorinlistoftodos,textwidth=\marginparwidth]{todonotes}

\title{Gradient Coding with Clustering and Multi-message Communication\\\thanks{This work was supported by EC H2020-MSCA-ITN-2015 project SCAVENGE under grant number 675891, and by the European Research Council project BEACON under grant number 677854.}}

\author{\IEEEauthorblockN{Emre Ozfatura\IEEEauthorrefmark{2},
Deniz G\"und\"uz\IEEEauthorrefmark{2} and Sennur Ulukus\IEEEauthorrefmark{3}}\\
\IEEEauthorblockA{ \IEEEauthorrefmark{2}Information Processing and Communications Lab, Dept. of Electrical and Electronic Engineering,\\ Imperial College London \\
\IEEEauthorrefmark{3}Department of Electrical and Computer Engineering, Institute for Systems Research,\\ University of Maryland, College Park \\
        {\tt \{m.ozfatura,d.gunduz\}@imperial.ac.uk, ulukus@umd.edu}
}}

\begin{document}
\maketitle
\begin{abstract}
Gradient descent (GD) methods are commonly employed in machine learning problems to optimize the parameters of the model in an iterative fashion. For problems with massive datasets, computations are distributed to many parallel computing servers (i.e., \textit{workers}) to speed up GD iterations. While distributed computing can increase the computation speed significantly, the per-iteration completion time is limited by the slowest \textit{straggling} workers. Coded distributed computing can mitigate straggling workers by introducing redundant computations; however, existing coded computing schemes are mainly designed against persistent stragglers, and partial computations at straggling workers are discarded, leading to wasted computational capacity. In this paper, we propose a novel gradient coding (GC) scheme which allows multiple coded computations to be conveyed from each worker to the master per iteration. We numerically show that the proposed GC with multi-message communication (MMC) together with clustering provides significant improvements in the average completion time (of each iteration), with minimal or no increase in the communication load.   
\end{abstract}

\section{Introduction}

In centralized synchronous distributed gradient descent (DGD), each worker computes a gradient estimate, called a \textit{partial gradient}, using its own local dataset, which are then aggregated by the parameter server, i.e., the \textit{master}, to update the model parameters \cite{PSGD1}. In this implementation, the master waits until it receives the partial gradients from all the workers; hence, the overall computation speed  is limited by the slowest, so-called \textit{straggling} workers. Straggler tolerant DGD schemes have been introduced recently, where redundant computations are assigned to workers, either together with coded dataset \cite{CC.1,CC.3,CC.4, CC.6, CC.7, CC.8, CC.11}, or combined with coded local computations \cite{UCCT.1,UCCT.2,UCCT.3}, or by simply using backup computations \cite{UCUT.2,UCUT.4,UCUT.5}.
 
Most of these works target \textit{persistent stragglers}; that is, the straggling workers either cannot fulfill any computations, or their computations are discarded; hence, the computation capacity of the network is underutilized. However, in practice, we mostly encounter {\em non-persistent stragglers}, which are not capable of completing all their assigned tasks by a deadline, but may still compute a significant portion of them. Partial computations carried out by non-persistent stragglers can also be utilized by the master to reduce the per-iteration completion time \cite{CC.9, CC.14,UCUT.4,UCUT.2,CC.13}. 

In this paper, we introduce a {\em multi-message communication (MMC)} strategy for centralized synchronous DGD to better utilize the computational capacity and to reduce the computation time. In particular, we extend the gradient coding (GC) approach of \cite{UCCT.1,UCCT.3} to MMC in order to seek a balance between the communication load and the computation time. MMC has been previously considered with different straggler avoidance strategies \cite{CC.2,CC.9, CC.14,CC.5,UCUT.4,UCUT.2,CC.13}. Nevertheless, most of these studies are either limited to linear regression type problems, where the gradient computation boils down to matrix-matrix or matrix-vector multiplication \cite{CC.2,CC.9, CC.14,CC.5,CC.13}, or simply assign the same computation to multiple workers \cite{UCUT.4,UCUT.2}, which leads to high communication load. Hence, the aim of the proposed design is to reduce the computation time with MMC while avoiding high communication load, unlike the aforementioned uncoded repetition schemes, and to ensure that the design is applicable to a wide range of machine learning problems employing DGD. In this paper, we achieve these objectives by introducing a novel GC scheme, called GC-MMC, in which multiple coded partial gradients are generated and transmitted by each worker as computations go on. This allows the master to exploit the partial computations of straggling workers, which is particularly useful when the workers have similar speeds (i.e., non-persistent stragglers), and also allows early termination before the straggling threshold is met, even in the presence of persistent stragglers. The latter is beneficial when a few workers are significantly faster than the others. Assuming a shifted exponential computation time statistics, we numerically show the benefits of the proposed GC-MMC scheme compared to other coded and uncoded computation schemes with or without MMC.

\section{Gradient Coding (GC)}
% In many machine learning problems, for given $N$ training data points $\mathbf{X}=[\mathbf{x}_{1},\ldots,\mathbf{x}_{N}]^{T}$, $\mathbf{x}_i \in \mathbb{R}^{d}$, and the corresponding labels $\mathbf{y}=[y_{1},\ldots,y_{N}]^{T}$,  $y_{i}\in \mathbb{R}$, $i \in[N]\defeq \{1,2,\ldots,N\}$, the objective is to minimize the {\em parameterized empirical loss function}
%\begin{equation}
%L(\boldsymbol{\theta}) \triangleq \sum_{i=1}^{N}l\left((\mathbf{x}_{i}, y_{i}),\boldsymbol{\theta} \right) + \lambda R(\boldsymbol{\theta}),
%\end{equation}
%where $\boldsymbol{\theta}\in \mathbb{R}^{d}$ is the parameter vector, $l$ is an application specific loss function, and $R(\boldsymbol{\theta})$ is the regularization component. This optimization problem is commonly solved by gradient descent (GD), where at each iteration, the parameter vector $\boldsymbol{\theta}\in\mathbb{R}^{d}$ is updated along the GD direction:
%\begin{equation}\label{update}
%\boldsymbol{\theta}_{t+1} = \boldsymbol{\theta}_{t} - \eta_{t} \nabla_{\boldsymbol{\theta}} L(\boldsymbol{\theta}), ~
%\end{equation}
%where $\eta_{t}$ is the learning rate at iteration $t$,
%and the gradient at the current parameter vector is given by $\nabla_{\boldsymbol{\theta}}=\sum_{i=1}^{N}\nabla_{\boldsymbol{\theta}}l\left((y_{i},x_{i}),\boldsymbol{\theta})\right)$.

Consider DGD implemented with one master and $K$ workers. The dataset $\mathbf{X}$ and the corresponding labels $\mathbf{y}$ are divided into $K$ non-overlapping equal-size mini-batches $B_{1},\ldots,B_{K}$. We use  $g^{(t)}_{k}$ to denote the partial gradient for parameter vector $\boldsymbol{\theta}_{t}$ evaluated with mini-batch $B_{k}$, i.e.,
\begin{equation}
g^{(t)}_{k}=\frac{1}{\vert B_{k} \vert} \sum_{(\mathbf{x}_{i}, y_{i})\in B_{k}} \nabla l((\mathbf{x}_{i}, y_{i}),\boldsymbol{\theta}_{t}).
\end{equation}
%\footnote{In \cite{UCCT.1}, two schemes are introduced namely; fractional repetition and cyclic repetition. In this paper, by GC we refer to the cyclic repetition scheme.}
The ``full gradient'' computed over the whole  dataset, can be obtained by $\frac{1}{K}\sum^{K}_{i=1}g^{(t)}_{i}$. GC is a distributed computing framework to compute the full gradient across $K$ workers while tolerating straggling workers thanks to redundant computations assigned to each worker \cite{UCCT.1} 
%consists of three steps: i) \textit{mini-batch assignment}, where multiple mini-batches are assigned to each worker for redundancy; ii) \textit{gradient coding}; and iii) \textit{gradient encoding}. 
Let $\mathcal{B}_{k}$ denote the set of mini-batches assigned to the $k$th worker, i.e., if $B_{i}\in\mathcal{B}_{k}$, then $g^{(t)}_{i}$ is computed by the $i$th worker. The \textit{computation load}, $r$, is defined as the number of mini-batches assigned to each worker, i.e., $\vert\mathcal{B}_{k}\vert = r$, $\forall k\in[K]$. After computing $r$ partial gradients, one for each mini-batch available locally, each worker sends a linear combination of those computations, $c^{(t)}_{k}\defeq\mathcal{L}_{k}(g^{(t)}_{i}:B_{i}\in\mathcal{B}_{k})$, called the {\em coded  partial gradient}. The master waits until it receives sufficiently many coded partial gradients to obtain the full gradient. It is shown in \cite{UCCT.1} that, for any set of non-straggler workers $\mathcal{W}\subseteq[K]$ with  $ \lvert \mathcal{W}   \rvert = K-r+1 $, there exists a set of coefficients $\mathcal{A}_{\mathcal{W}}=\left \{a^{(t)}_{k}:k\in\mathcal{W}\right\}$ such that $\sum_{k\in\mathcal{W}} a^{(t)}_{k}c^{(t)}_{k}=1/K\sum^{K}_{k=1}g^{(t)}_{k}.$
Hence, GC can tolerate up to $r-1$ persistent stragglers at each iteration $t$. Next, we introduce two strategies, namely \textit{MMC} and \textit{clustering} to reduce the per-iteration completion time with GC.

\section{Gradient Coding with Multi-Message Communication (GC-MMC)}

In the original GC scheme of \cite{UCCT.1}, the number of messages transmitted to the master per-iteration per-worker is limited to one. Due to the synchronized model update, the workers that complete their computations stay idle until they receive the updated parameter vector to start the next iteration.  Besides, since the workers send a single message only after completing all the assigned computations, computations carried out by the  slow/straggling workers are discarded. To prevent the under-utilization of the computation resources, we will allow each worker to send coded partial gradients to the master; that is, at each iteration each worker sends multiple coded partial gradients instead of sending a single computation result. In the scope of this paper, we will present two different approaches to design coded partial gradients, namely {\em correlated code design} and {\em uncorrelated code design}, which are explained next.

\subsection{Correlated code design}

In GC, the number of partial gradients linearly combined to form the transmitted message from a worker is equal to the computation load $r$. In MMC, we allow each worker to compute and transmit multiple coded partial gradients, each of which will be generated by combining $m \leq r$ gradient computations. We will refer to $m$ as the {\em order} of the corresponding partial gradient. In particular, each worker will be able to send up to $l= r-m+1$ different messages, each of order $m$; that is, each of the coded partial gradients will be a linear combination of the $m$ most recently computed partial gradients. 

As an example, let $r=3$ and $m=2$, and consider the worker with mini-batches $B_{1}, B_{2}$ and $B_{3}$, which will compute partial gradients $g_{1}, g_{2}$ and $g_{3}$ in this order\footnote{In the rest of the paper, to simplify the notation we will drop the time index from the gradients when we focus on a single iteration of the algorithm.}. After computing $g_{1}$ and $g_{2}$ the worker will send a linear combination of these two partial gradients as a coded message to the master, and after computing $g_{3}$, it will send a linear combination of partial gradients $g_{2}$ and $g_{3}$.

In general, the proposed scheme consists of two steps:  coded message construction and message assignment. In the coded message construction step, structure of the coded messages are designed according to the order $m$ by simply setting $r=m$ in the original GC scheme. Then, in the coded message assignment step, constructed messages are assigned to the workers based on the assigned partial gradients. We present the following example to clarify these steps.

\textbf{Example 1:} Let $K=6$, $r=3$, $m=2$, and consider the assignment matrix $\mathbf{M}$, whose $i$th row indicates the mini-batches assigned to the $i$th worker; that is  $\mathbf{M}(i,j)=1$ means that partial gradient $g_{j}$ will be computed by the $i$th worker. In GC with $K=6$ and $r=3$, we have the following assignment matrix.
\begin{equation}
\mathbf{M}=
  \begin{bmatrix}
    1 & 1 & {\color{red}1} & 0 & 0 & 0  \\
    0 & 1 & 1 & {\color{red}1} & 0 & 0 \\
    0 & 0 & 1 & 1 & {\color{red}1} & 0 \\
    0 & 0 & 0 & 1 & 1 & {\color{red}1} \\
    {\color{red}1} & 0 & 0 & 0 & 1 & 1\\
    1 & {\color{red}1} & 0 & 0 & 0 & 1 \\
  \end{bmatrix}
\end{equation}
When  $m=2$, coded gradients are obtained according to the assignment matrix $\tilde{M}$, which is obtained by removing the last $r-m$ of the 1s in each row (shown in {\color{red}red} above). When the assignment matrix $\tilde{M}$ is used to design GC, a total of $K=6$ coded partial gradients, each of order two, are constructed; and the full gradient can be obtained from any $K-m-1=5$ coded partial gradients. Let ${c_{1},\ldots,c_{6}}$ denote the corresponding coded partial gradients. We remark that $c_{1}$ is a  linear combination of $g_{1}$ and $g_{2}$, while $c_{2}$ is a linear combination of $g_{2}$ and $g_{3}$. Since $g_{1},g_{2}$ and $g_{3}$ are assigned to the first worker, it can send both coded messages $c_{1}$ and $c_{2}$. 
%To illustrate the assignment of coded partial gradients, we use the assignment matrix $\mathbf{C}$, where the $i$th column shows the assigned coded gradients to the $i$th worker in the order of computation. For, Example 1, we have
%\begin{equation}
%\mathbf{C}=
%  \begin{bmatrix}
%%   c_{1} & c_{2} & c_{3} & c_{4} & c_{5} & c_{6} \\
%    c_{2} & c_{3} & c_{4} & c_{5} & c_{6} & c_{1} \\
%  \end{bmatrix}.
%\end{equation}
We call this approach {\em correlated code design}, since the same coded partial gradient can be computed and sent by more than one worker, e.g., in Example 1, $c_{2}$ can be sent by both worker 1 and worker 2. In Example 1, the original GC algorithm needs to receive computations from at least four workers in order to complete an iteration; whereas the proposed scheme can complete an iteration with results from only three workers. For instance, when workers 1, 2 and 4 each send two coded partial gradients, the master will obtain $c_{1},c_{2},c_{3},c_{4},c_{5}$, and recover the full gradient. In the next section, we will analyze the uncorrelated coded design approach, where each coded gradient is assigned to exactly one worker.

\subsection{Uncorrelated code design}

As explained in \cite{UCCT.3}, GC can be interpreted as a polynomial interpolation problem. In this model, the gradient assignment matrix is called  the {\em mask matrix}, and the {\em support} of the $i$th row $\mathbf{M}_{(i \text{ } , \text{ }:)}$, denoted by $supp(\mathbf{M}_{(i \text{ } , \text{ }:)})$, gives the index of the partial gradients that are assigned to the $i$th worker. For a given mask matrix $\mathbf{M}$, GC is equivalent to interpolating a polynomial with {\em degree} $h$, where $h=N-\max_{i}{\vert supp(\mathbf{M}_{(: \text{ } , \text{ }i)})\vert}$; in other words, $h$ is equal to the number of zeros in the most sparse column of $\mathbf{M}$. In a broad sense, each partial gradient $g_{k}$ is embedded into a polynomial $f_{k}$, and each worker evaluates the  polynomials $f_{1},\ldots,f_{K}$ at preassigned points, and sends their sum to the master. The key design issue is that, if gradient $g_{k}$ is not assigned to the $j$th user, then the value of polynomial $f_{k}$ should be zero at the point assigned to that user. Hence, if partial gradient $g_{k}$ is missing at $h$ workers, which is the number of zeros in the $j$th column of the mask matrix $\mathbf{M}$, then the corresponding polynomial $f_{k}$ should have $h$ roots, which implies that $f_{k}$, and hence, the polynomial $\sum_{k}f_{k}$, both have degree $h$.

Remember that a polynomial with degree $h$ can be interpolated from its values at $h+1$ different points. Hence, as in the coded gradient approach in \cite{UCCT.3} non-straggling threshold $K_{th}=K-r+1$ can be achieved utilizing a Reed-Solomon code structure. Please refer to \cite{UCCT.3} for the details\footnote{Implementation of the encoding and decoding procedures also effect the completion time; however, this is ignored in the scope of this paper.}.\\
\indent In \textbf{Example 1}, the mask matrix $\mathbf{M}$ has exactly 3 zeros in each column; therefore, at least 4 coded partial gradients must be received to recover the full gradient. As in GC, the main drawback of the scheme in \cite{UCCT.3} is that the workers send coded partial gradients only after computing all the assigned partial gradients. We will design a GC scheme with MMC, in which each coded partial gradient is sent by at most one worker. If each worker is allowed to send $l$  messages per iteration, we consider a setting with $Kl$ workers, introducing $K (l-1)$ ``virtual'' workers. Then, we design a GC scheme for the mask matrix of $Kl$ users. We explain the proposed MMC strategy and the concept of virtual workers on Example 1.

\textbf{Example 1 (continued):} Since each worker can send $l =2$ coded partial gradients, we create one virtual worker corresponding to each real worker. Partial gradient computations assigned to a virtual worker should be a subset of those assigned to the corresponding real worker. Since $r=3$  partial gradients are assigned to each worker, we assign the first two partial gradients of each real worker to the corresponding virtual worker. The corresponding mask matrix for all the workers (including the virtual workers) is given below:
\begin{equation}
\mathbf{M}=
  \begin{bmatrix}
    {\color{red} 1} & {\color{red}1} & {\color{red}1} & {\color{red}0} & {\color{red}0} & {\color{red}0}  \\
    {\color{blue} 1} & {\color{blue}1} & {\color{blue}0} & {\color{blue}0} & {\color{blue}0} & {\color{blue}0}  \\
    {\color{red}0} & {\color{red}1} & {\color{red}1} & {\color{red}1} & {\color{red}0} & {\color{red}0} \\
        {\color{blue} 0} & {\color{blue}1} & {\color{blue}1} & {\color{blue}0} & {\color{blue}0} & {\color{blue}0}  \\
    {\color{red}0} & {\color{red}0} & {\color{red}1} & {\color{red}1} & {\color{red}1} & {\color{red}0} \\
        {\color{blue} 0} & {\color{blue}0} & {\color{blue}1} & {\color{blue}1} & {\color{blue}0} & {\color{blue}0}  \\
    {\color{red}0} & {\color{red}0} & {\color{red}0} & {\color{red}1} & {\color{red}1} & {\color{red}1} \\
        {\color{blue} 0} & {\color{blue}0} & {\color{blue}0} & {\color{blue}1} & {\color{blue}1} & {\color{blue}0}  \\
    {\color{red}1} & {\color{red}0} & {\color{red}0} & {\color{red}0} & {\color{red}1} & {\color{red}1} \\
        {\color{blue} 0} & {\color{blue}0} & {\color{blue}0} & {\color{blue}0} & {\color{blue}1} & {\color{blue}1}  \\
    {\color{red}1} & {\color{red}1} & {\color{red}0} & {\color{red}0} & {\color{red}0} & {\color{red}1} \\
        {\color{blue} 1} & {\color{blue}0} & {\color{blue}0} & {\color{blue}0} & {\color{blue}0} & {\color{blue}1}  \\
  \end{bmatrix},
\end{equation}
where the rows in {\color{red}red} correspond to the real workers, while the rows in {\color{blue}blue} to the virtual workers. One can easily observe that in each column there are exactly $7$ zeros; and hence, the master needs at least $8$ coded gradients to recover the full gradient.

We remind that, in the original GC scheme, the master waits for $4$ workers to recover the full gradient. However, in the proposed scheme each worker can send a message with a coded partial gradient as a virtual worker, after computing only two partial gradients, and the full gradient can be recovered from any 8 coded partial gradients, including those from the virtual workers. Consider, for example, three of the workers are non-straggler, and each of them sends 2 coded partial gradients, while two workers are non-persistent stragglers, each of them sends one coded gradient, and the last worker is a persistent straggler. In this case the full gradient can be obtained by the proposed approach but not with the original GC scheme. Hence, the proposed approach improves the per-iteration completion time.

Uncorrelated code design for GC with MMC is defined by the order vector $[m_{0},\ldots,m_{i}]$, where $m_{0}$ is the order of the coded partial gradient sent by the real worker, and $m_{i}$ denotes the order of the coded partial gradient sent by the $i$th virtual worker. For a particular MMC strategy with order vector $\mathbf{m} = [m_{0},\ldots,m_{i}]$, the number of zeros in any column of the mask matrix $M$ is given by $Kl-(\sum^{l}_{i=0}m_{i})$; and thus,  $Kl-(\sum^{l}_{i=0}m_{i})+1$  coded partial gradients are required to recover the full gradient. 

We emphasize here that the use of coded partial gradients with lower orders increases the number of messages required for recovering the full gradient; however, they can be obtained faster; and hence, we can exploit the computations carried out by non-persistent stragglers. The focus of the current paper is on identifying potential designs for GC with MMC; thus, the optimization of the partial gradient orders is left as future work.

Another important issue regarding MMC is the \textit{communication load}, which denotes the average number of messages received by the master at each iteration. The communication load increases with the number of virtual workers; therefore, the optimal MMC strategy  depends critically on the communication architecture of the network and the protocol used to transmit messages from the workers to the master as well as the computation speeds of the workers. 

\subsection{Clustering}

Next we introduce clustering, which can further speed up the computation time. Let the workers be divided into $P$ equal-size disjoint clusters, where the set of workers in cluster $p$ is denoted by $\mathcal{K}_{p} \subset [K]$, $p\in[P]$. Mini-batches are also divided into equal-size disjoint subsets, and the set of mini-batches assigned to $p$th cluster is denoted by $\mathcal{B}_{\mathcal{K}_{p}}$. In the clustering approach, the workers in the $p$th cluster are responsible for computing
\begin{equation}
\frac{1}{\vert\mathcal{B}_{\mathcal{K}_{p}}\vert}\sum_{k:B_k \in \mathcal{B}_{\mathcal{K}_{p}} }g^{(t)}_{k},
\end{equation}
and the GC scheme is applied to each cluster independently.

At this point, we remark that the fractional repetition scheme in \cite{UCCT.1} is a special case of the clustering approach where the size of a cluster is equal to $r$. Consider, $K=40$, $r=10$ and $P=4$, where the workers are divided into $P=4$ clusters, while the mini-batches are divided into $4$ subsets, and each cluster is responsible for a different subset. In the fractional repetition scheme the master waits until at least one worker from each cluster completes and sends its partial gradient. One can observe that if there are $K-r+1=31$ workers that have completed and sent their computations to the master, there must be at least one worker from each cluster among these workers; hence, the non-straggling threshold is  $K-r+1$. However, the non-straggling threshold represents a worst case scenario. Notice that, even 4 workers, each from a different cluster can be sufficient to obtain the full gradient. On the other hand, the other GC scheme proposed in \cite{UCCT.1}, called the cyclic repetition scheme, always has to wait until receiving coded messages from $K-r+1$ worker. Hence, we can state that although various GC scheme can achieve the same optimal non-straggling threshold, their average performance may differ. 

While the fractional repetition scheme  requires $K$ to be an integer multiple of $r$, the clustering approach outlined above can be applied to any $(K,r)$ pair. When GC is applied with clustering, it is possible to tolerate $r-1$ stragglers in each cluster; thus, for a particular straggler realization, if the full gradient can be obtained with GC (without clustering) then it can also be obtained with clustering, while the converse is not true. Based on this observation, we can claim that with clustering, straggling threshold remains the same, but the average computation time can be reduced. However, when MMC is allowed, clustering may also be disadvantageous. On one hand, more straggling workers can be tolerated on average, on the other hand  GC is applied to each cluster independently; hence, a coded partial gradient from a particular cluster cannot be utilized for another cluster. Consequently, the optimal clustering strategy with MMC depends on the computation statistics of the workers.

\subsection{Hybrid implementation}
The optimal DGD strategy depends critically on the computation time statistics of the workers. In particular, when the computation speeds of the workers are similar, MMC is expected to have a better performance as it can exploit all the computations carried out across the workers; however, when one of the workers is much faster compared to the others, fractional repetition can be preferred against MMC. To illustrate this trade-off, consider the case $K=5$ and $r=5$. With the fractional repetition scheme, the master waits for the fastest worker to finish all assigned computations; however, with GC-MMC with order vector $\mathbf{m}= [5,3]$, the master waits for exactly 3 coded messages. Accordingly, we can propose a hybrid scheme, in which the workers initially behave as dictated by the GC-MMC scheme, but if a worker is fast enough to complete all its computations, then it switches to fractional repetition scheme, and sends the average gradient instead of a coded partial gradient. 

\begin{figure*}
     \centering
          \begin{subfigure}[b]{0.47\textwidth}
         \includegraphics[scale=0.5]{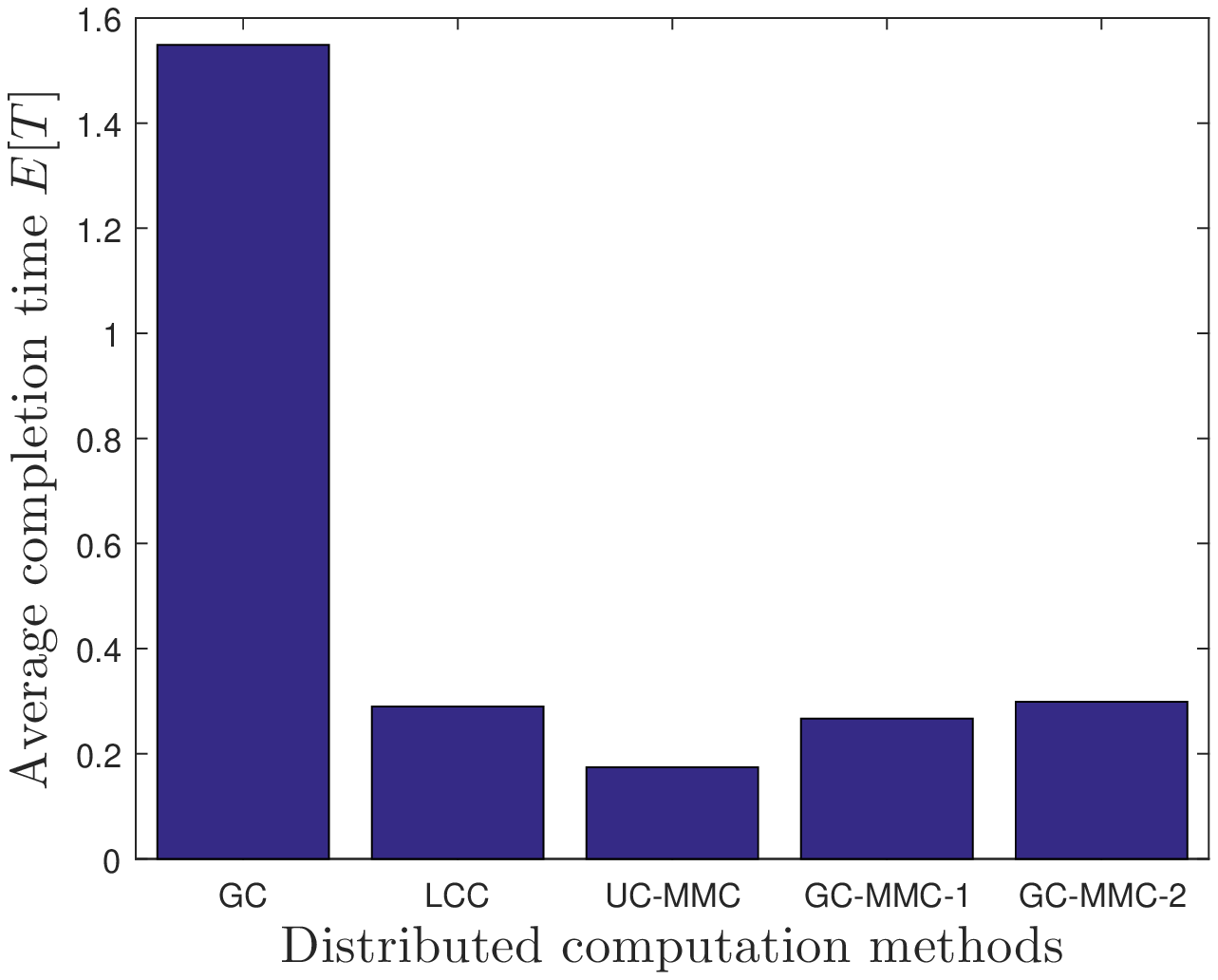}
        \caption{Average completion time}
 				\label{comp}
     \end{subfigure}
    \begin{subfigure}[b]{0.47\textwidth}
        \includegraphics[scale=0.5]{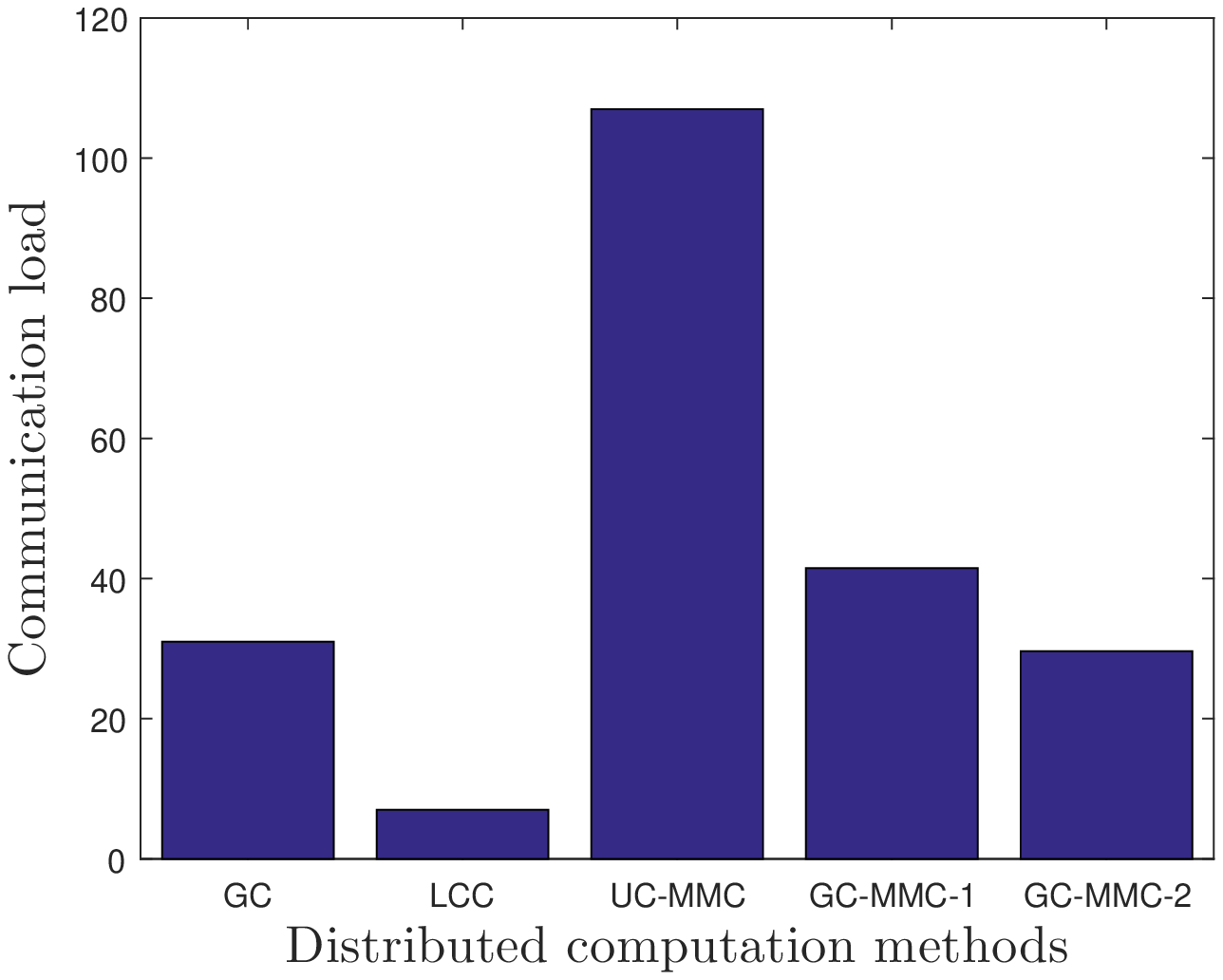}
         \caption{Communication load}
 				\label{comm}
         \end{subfigure}
 				\caption{Average per-iteration completion time and communication load under the shifted-exponential model.}
 		\label{avg_perf}
 \end{figure*}

\section{Numerical Results and Discussions}

\subsection{Simulation Results}

For the simulations, we consider the setup in \cite{CC.4}, where there are $K=40$ workers and the dataset is divided into $K=40$ mini-batches. We set $r=10$; that is, 10 partial gradient computations, each corresponding to a different mini-batch, are assigned to  each worker.
For the computation time statistics of the workers, we adopt the commonly used shifted exponential model \cite{CC.1,CC.2}, which assumes that the probability of completing $s$ computations (partial gradients) at any worker by time $t$ is given by
\begin{equation}\label{dist}
F_{s}(t)\defeq
    \begin{cases}
     1-e^{-\mu(\frac{t}{s}-\alpha)}, &  \text{if } t\geq s\alpha, \\
      0,   &  else. 
    \end{cases}
\end{equation}
In our simulations,  we used parameters $\mu=10$ and $\alpha=0.01$, which implies that, on average, $\mu=10$ partial gradients can be computed by a worker in unit time (e.g., one second), and the computation time of a single partial gradient cannot be less than $\alpha=0.01$.\\
\indent For the performance analysis we design two different GC schemes with MMC. In the first scheme, called GC-MMC-1, four equal-size clusters are formed and correlated code design is employed in each cluster with order $m=6$, so that each worker can send up to 5 coded partial gradients. In the second scheme, called GC-MMC-2, again four equal-size clusters are formed, but uncorrelated code design is employed in each cluster with order vector $\mathbf{m}=[6, 8, 10]$. We compare the performance of these schemes with GC, uncoded computation with MMC (UC-MMC) \cite{CC.5} and Langrange coded computation (LCC) \cite{CC.4} schemes, in terms of the average per-iteration completion time and the average communication load, i.e., the average number of messages sent to the master at each iteration.

The results are presented in Figure \ref{avg_perf}. We observe that the
GC-MMC schemes can perform as well as LCC in terms of the average per-iteration completion time, although LCC introduces significant coding and decoding complexity. Furthermore, the GC-MMC schemes can even outperform the LCC if the number of messages are increased. Indeed, notice that the UC-MMC scheme is actually a special case of the GC-MMC scheme with correlated code design with order $m=1$. However, this improvement in the average per-iteration completion time performance comes at the expense of an increase in the communication load. The key advantage of the GC-MMC schemes is the more flexible trade-off they provide between the completion time and the communication load compared to UC-MMC and GC. To this end, an interesting observation is that the GC-MMC-2 achieves the same performance as LCC, a coded computation scheme, while still sending only a single message per-worker per-iteration on average.

We want to highlight that LCC, its multi-message variations \cite{CC.5} might be more efficient compared to GC-MMC schemes in terms of the completion time as well as the communication load statistics. However, LCC and other coded computation schemes have the following drawbacks in real applications i) they require encoding the whole dataset before computation; ii) they are limited to problems where the gradient computation can be reduced to matrix-vector or matrix-matrix multiplication (mainly linear regression problems);  iii) other techniques that can increase the communication efficiency, such as gradient quantization \cite{SGD_q1,SGD_q5_com} cannot be incorporated with them. Therefore, we believe that speeding up coded computation schemes that compute uncoded data points is an important research direction for practical applications. 
\section{Conclusions}
We introduced a novel GC scheme with MMC and clustering, allowing workers to send coded partial gradients before completing all of their assigned computations. This can potentially lead to early termination of an iteration as it can benefit from computations carried out at all the workers, including the relatively slower stragglers. The improvement is achieved at the expense of an increased communication load, and the proposed scheme seeks a balance between the communication load and the average per-iteration completion time, thanks to its flexible coding and clustering structure. We numerically show that, under certain completion time statistics, average completion time can be reduced dramatically with a very small or no increase in the communication load. As a future extension, we will design a partially recoverable GC scheme, similarly to \cite{CC.14}, in order to provide further flexibility to the master.
\bibliographystyle{IEEEtran}
\bibliography{IEEEabrv,ref}
\end{document}